\documentclass[floatfix,aps,nofootinbib,prd,twocolumn,groupedaddress,superscriptaddress]{revtex4}
\usepackage{amsmath}
\usepackage{dcolumn}
\usepackage{float}
\usepackage{physics}
\usepackage{appendix}
\def\cat@comma@active{\catcode`\,12}%
\usepackage{breqn}
\usepackage{graphicx}
\usepackage[pdftex]{hyperref}
\usepackage[dvipsnames]{xcolor}
\usepackage{comment}
\usepackage[utf8]{inputenc}
\usepackage{cases}
\newcommand{\be}{\begin{eqnarray}}
\newcommand{\ee}{\end{eqnarray}}
\newcommand{\e}{\mathrm{e}}
\newcommand{\dmdt}{-\partial_T \langle m \rangle}

\makeatletter
\let\cat@comma@active\@empty
\makeatother
\begin{document}
\title{Moving from continuous to discrete symmetry in the 2D XY model}
\author{Nouman Butt}
\author{Xiao-Yong Jin}
\author{James C. Osborn}
\affiliation{Computational Science Division,  Argonne National Laboratory,  9700 S Cass Ave, Lemont IL 60439}
\author{Zain H. Saleem}
\affiliation{Mathematics and Computer Science Division, 
Argonne National Laboratory, 9700 S Cass Ave, Lemont IL 60439}
\date{\today}
\begin{abstract}
We study the effects of discretization on the U(1) symmetric XY model in
two dimensions using the
Higher Order Tensor Renormalization Group (HOTRG) approach.
Regarding the $Z_N$ symmetric clock models as specific discretizations
of the XY model,
we compare those discretizations to ones from truncations of the tensor network
formulation of the XY model based on a character expansion,
and focus on the differences in their phase structure at low temperatures.
We also divide the tensor network formulations into core and interaction tensors
and show that the core tensor has the dominant influence on the phase structure.
Lastly, we examine a perturbed form of the XY model that continuously interpolates
between the XY and clock models.
We examine the behavior of the additional phase transition caused by the perturbation
as the magnitude of perturbation is taken to zero.
We find that this additional transition has a non-zero critical
temperature as the perturbation vanishes, suggesting
that even small perturbations can have a significant effect on the
phase structure of the theory.
\end{abstract}

\maketitle

\section{Introduction}

While quantum computers have the potential to allow great advances in the
simulation of quantum systems, a major challenge will be making efficient
use of the limited resources available in near-term hardware.
For systems with continuous degrees of freedom, such as in common
gauge theories, representing the continuous field
in a limited set of discrete states poses a challenge.
For quantum Hamiltonians, the mapping to an infinite set of discrete states
can be done through second quantization,
but the set of states then needs to be truncated to fit on a quantum computer.
There have been several studies of methods for discretizing and truncating
field representations in lattice models (both classical and quantum),
including spin models \cite{PhysRevD.93.085012,PhysRevLett.123.090501,PhysRevD.100.054505,
PhysRevD.99.074501,PhysRevLett.126.172001,PhysRevB.103.245137},
scalar fields \cite{doi:10.1126/science.1217069,10.5555/2685155.2685163,
PhysRevA.98.042312,PhysRevA.99.052335},
and gauge theories \cite{BROWER2004149,PhysRevA.73.022328,PhysRevD.91.054506,
PhysRevD.95.094509,PhysRevX.7.041046,PhysRevA.99.062341,PhysRevD.100.114501,
PhysRevD.102.114513,PhysRevD.102.114517,Haase2021resourceefficient,PhysRevD.103.094501}
.

A similar truncation occurs when simulating these theories using tensor networks 
\cite{Liu:2013nsa}.
Here we use this correspondence to study the effects of truncation in two dimensional
classical spin models as a proxy for mapping models to a quantum computer.
The example systems we will use are the $XY$ model and it's discrete analog,
the $Z_N$ symmetric clock model.
The two-dimensional XY model has a continuous $U(1)$ symmetry and exhibits a single phase
transition of the Berezinskii–Kosterlitz–Thouless (BKT) type
\cite{1971JETP...32..493B,1972JETP...34..610B,Kosterlitz_1973}.
The simplest discretization of the XY model, using $N$ discrete states, produces the
$Z_N$ clock-models, which break the $U(1)$ symmetry.
This discretization method has a major effect on the phase structure of the theory.
While the XY model is recovered in the $N \to \infty$ limit, one may need to go to very
large $N$ before the discretization effects become negligible at low temperatures.

The $Z_N$ clock models have been studied extensively, and their phase diagrams are
known to be distinctly different from the XY model.
For $N \leq 4$, they exhibit a single phase transition, though due to
spontaneous symmetry breaking, and not of the BKT type.
For $N \geq 5$, there is clear evidence that the
models exhibit two phase transitions.
The transition at higher temperature becomes the BKT transition of the XY model as $N \to \infty$.
For sufficiently large, but finite $N$, it is also consistent with a BKT transition,
even though there is only a discrete symmetry.
This is an example of an emergent symmetry.
An explanation for the emergent symmetry in the clock models was provided based
on a mapping of the classical model to a quantum Hamiltonian \cite{Sun:2019ryo,Ortiz:2012zz},
which also suggested that the upper transition is BKT for all $N\ge 5$.
However, numerical studies suggest that the $N=5$ transition may not be 
the same type as for $N>5$, though may still be related 
\cite{PhysRevE.82.031102,PhysRevE.83.041120,PhysRevE.88.012125}.
The nature of the lower temperature transition is also believed to be of BKT-type based
on central charge arguments
\cite{PhysRevLett.56.742,PhysRevLett.56.746,PhysRevB.101.165123}.
While the upper critical temperature remains fairly constant as $N \to \infty$,
the lower critical temperature moves towards zero in that limit \cite{PhysRevE.98.032109}.
It is the existence of the lower critical temperature that largely distinguishes the
$Z_N$ models for $N\ge 5$ from the XY model.

Other discretization methods for the XY model are possible.
In particular, the formulation of the XY model as a tensor network,
which can be done using a character expansion, provides a convenient
basis for truncation, and is manifestly $U(1)$ invariant.
Similar tensor network constructions using character bases exist for gauge theories,
and have been proposed to be useful for mapping gauge theories to qubits while
preserving gauge invariance \cite{RevModPhys.94.025005}.

The differences between truncation schemes for the XY model can be
better understood by examining the differences in their tensor network formulations.
It was shown that the tensor network formulation of the XY model preserves
the U(1) symmetry through the enforcement of selection rules \cite{PhysRevD.100.014506,PhysRevD.102.014506}.
The discrete version of the XY model, the $N$-state clock model,
can be seen as a version of the XY model where the infinite
set of states are now folded into the finite $N$ states in a periodic manner.
The periodic nature of the $Z_N$ symmetry alters
the selection rule and modifies the symmetry of the theory.
To further explore this we split the tensor construction into a core tensor,
centered on sites, and an interaction matrix, which connects neighboring sites,
and consider theories which mix the core and interaction from the different models.
From this we find that the core tensor, which enforces the selection rule, is indeed
the dominant factor on the phase structure of the theory, as opposed to
the interaction matrix.

We also examine the behavior of the phase structure when moving between the XY and $Z_N$
models.
As mentioned above, one way to transition from the clock model to the
XY model is to take $N$ large.
An alternative path is to introduce a symmetry breaking term into the XY model, that explicitly breaks the symmetry down to $Z_N$.
This is equivalent to the model studied
analytically by Jos\'e, Kadanoff, Kirkpatrick, and Nelson (JKKN)
\cite{PhysRevB.16.1217,PhysRevB.17.1477}.
They concluded that the $U(1)$ symmetry is unstable at low temperatures in response to small
values of the symmetry breaking field and the system develops a second phase transition
similar to the clock models.
The perturbation has the effect of smoothly transforming the XY core tensor
into the $Z_N$ clock model core tensor by the introduction of periodic terms.
We are interested in studying the emergence of the phase transition at low temperatures
in the $XY$ model for small values of the symmetry breaking term.
Our results suggest that even small perturbations can have a significant effect
on the low temperature phase structure of the XY model.
This may require caution when considering simulations of truncated models
to ensure that even small perturbations, perhaps due to discretization effects
or even simulation errors, don't give rise to new phases not present in the original theory.

In section \ref{sec:TN} we present the tensor network representation of
the $XY$ model, its perturbed version and the $Z_N$ clock model,
and elaborate on the difference in the structure of the core tensor in each case.
In section \ref{sec:disc} we compare numerical results for different
truncations of $XY$ model based on the character expansion and
the $Z_N$ clock model.
In section \ref{sec:role} we show results of mixing the core tensor and
interaction matrix from different models, and how it effects
the phase structure.
In section \ref{sec:pert} we perform a detailed study of the behavior of the
lower temperature phase transition of the perturbed model
as the symmetry breaking field becomes smaller, to determine the
effects of small symmetry breaking on the phase diagram.

\section{Tensor network formulation of XY and $Z_N$  spin models}

\label{sec:TN}

The Hamiltonians for the XY and $Z_N$ clock models can
be written in the common form
\begin{equation}
    H = - \sum_{\langle ij\rangle} \cos(\theta_i -\theta_j)
    - h \sum_{i} \cos (\theta_i)
\end{equation}
where the first sum is over neighboring sites on 
a 2D periodic lattice and $h$ is the magnetic field.
The angles, $\theta_i$, take continuous values in $[0,2\pi)$ for
the XY model and the discrete values $2\pi k/N$ with
$k\in{0,\ldots,N-1}$ for the $N$-state clock model.

The partition function for either model can be formulated as a tensor network by separating the interaction term through an expansion
in some basis, then integrating (summing for $Z_N$) over the
spin variables at each site \cite{Liu:2013nsa,Chen_2017}.
Details of this approach are given in Appendix \ref{sec:TNconstruction}.
The expansion produces an interaction matrix on the links of the lattice
while the integration produces a core tensor on the sites of the lattice.
The interaction matrix can then be factored and absorbed into the core
tensor to produce a single tensor per site which describes the partition function
of the system.

When defining the core tensor and interaction matrix, there is a choice of where
to put the local (single-site) terms in the Hamiltonian.
One choice is to keep the interaction matrix diagonal, which in the models considered here,
coincides with keeping only the interaction term 
$H_I(\theta_i,\theta_j) = - \cos(\theta_i - \theta_j)$
in the interaction matrix.
The choice of where to put the local terms has no effect on the partition function or
other physical results, and is only relevant when studying the roles of the core tensor
and interaction matrices on the phase structure as done in section \ref{sec:role}.

A common basis to use for the expansion of the interaction matrix
is the character basis, which, for the $U(1)$ symmetric XY model
is just the phases $\exp(i n \theta)$.
For the XY model, this gives the core tensor and interaction matrix
at inverse temperature $\beta$ of
\be
\label{CXY}
C^{XY}_{abcd} &=& I_{a-b+c-d}(\beta h) \\
\label{IXY}
M^{XY}_{ab} &=& I_a(\beta) \delta_{ab}
\ee
where $I_n(x)$ is a modified Bessel function.
The order of indices on the core tensor, $C^{XY}_{abcd}$, is such that
$a$ and $b$ (likewise $c$ and $d$) correspond to opposite directions along the same dimension.
Each of the indices on the XY model core tensor extend from
$-\infty$ to $\infty$.
In practice this needs to be truncated.
Here we choose to truncate them symmetrically in the range $-S,\ldots,S$
giving a total of $D = 2S+1$ states for each index.
We will study the effect of this truncation for various values of $D$ below.

For the $Z_N$ clock model, keeping only the interaction term $H_I(\theta_i,\theta_j)$ in the interaction matrix,
gives the core tensor and interaction matrix
\be
\label{CNN}
C^{N}_{abcd} &=& \sum_{\ell=-\infty}^{\infty} I_{a-b+c-d+\ell N}(\beta h) \\
\label{INN}
M^N_{ab} &=& \sum_{k=-\infty}^{\infty} I_{a+kN}(\beta) \delta_{ab} ~.
\ee
The indices on the clock model tensor are all finite, taking on $N$ distinct values.
Note that the difference in the XY and $Z_N$ tensors is 
that the clock model has the same terms from the XY model, but they
are folded around the finite set of $N$ states.
This is a consequence of the periodic nature of the basis functions
with discrete angles
$\exp(i n \theta_k) = \exp(i n \theta_{k+N})$ for $\theta_k = 2 \pi k/N$
(see Appendix \ref{sec:TNconstruction} for details).

As detailed in Appendix \ref{sec:TNconstruction}, the clock model
tensors can also be constructed using an infinite number of states which 
provides convenient approach to directly compare with the XY model.
In this case the tensors are
\be
\label{CN}
C^{N\infty}_{abcd} &=& \sum_{\ell=-\infty}^{\infty}
 I_{a-b+c-d+\ell N}(\beta h) \\
\label{IN}
M^{N\infty}_{ab} &=& I_{a}(\beta) \delta_{ab} ~.
\ee
and the indices now extend from $-\infty$ to $\infty$.
This is just the interaction matrix from the XY model combined with the
clock model core tensor, extended to an infinite number of states
on each index.
This form of the clock model provides a smooth
interpolation between it and the XY model by adjusting
the terms in the core tensor, which is what the
perturbed XY model studied here will do.

For the perturbed $XY$ case (JKKN model) the following term
is added to the Hamiltonian
\be
\label{pxy}
\delta H = -h_{N} \sum_{i} \cos(N\theta_i)
\ee
which breaks the U(1) symmetry down to $Z_N$.
This additional term modifies the core tensor, giving
\be
\label{CPXY}
C^{XYN}_{abcd} &=& \sum_{\ell=-\infty}^{\infty} I_{a-b+c-d+N\ell}(\beta h) 
\frac{I_\ell(\beta h_N)}{I_0(\beta h_N)} ~.
\ee
The normalization is chosen such that as $h_N \to \infty$
this reproduces the $Z_N$ clock model core tensor,
while at $h_N = 0$ this is equivalent to the XY model.

Given the interaction matrix and core tensor for a particular model,
the site tensor can be constructed as in Appendix \ref{sec:TNconstruction}. 
By contracting a 2D lattice network of site tensors,
the partition function, $Z$, can be evaluated.
Observables can be obtained from derivatives of the partition function,
either numerically by finite differences of parameters,
or by replacing a site tensor with its derivative
(an impurity tensor \cite{PhysRevB.81.174411}).
We use the impurity tensor method to measure the magnetization.
For the temperature derivatives (used in the specific heat and cross-derivative)
we use a the numerical derivative obtained from a local polynomial fit.

The contraction of the tensors is performed numerically using
the Higher Order Tensor Renormalization Group (HOTRG) method~\cite{PhysRevB.86.045139}.
This combines pairs of neighboring site tensors
to form a new blocked site tensor.
As the size of a tensor index grows above some limit, $D_{cut}$,
we employ a Higher Order Singular Values Decomposition
(HOSVD) to truncate it back to $D_{cut}$ states.
This is distinct from the initial $D$ states
in the truncation of the XY model.
In all cases we perform 20 blocking steps in each direction 
giving a final volume of $V = 2^{20} \times 2^{20}$ sites.

\begin{figure*}[htb!]
 \centering
 \begin{minipage}{0.48\textwidth}
  \includegraphics[width=\textwidth]{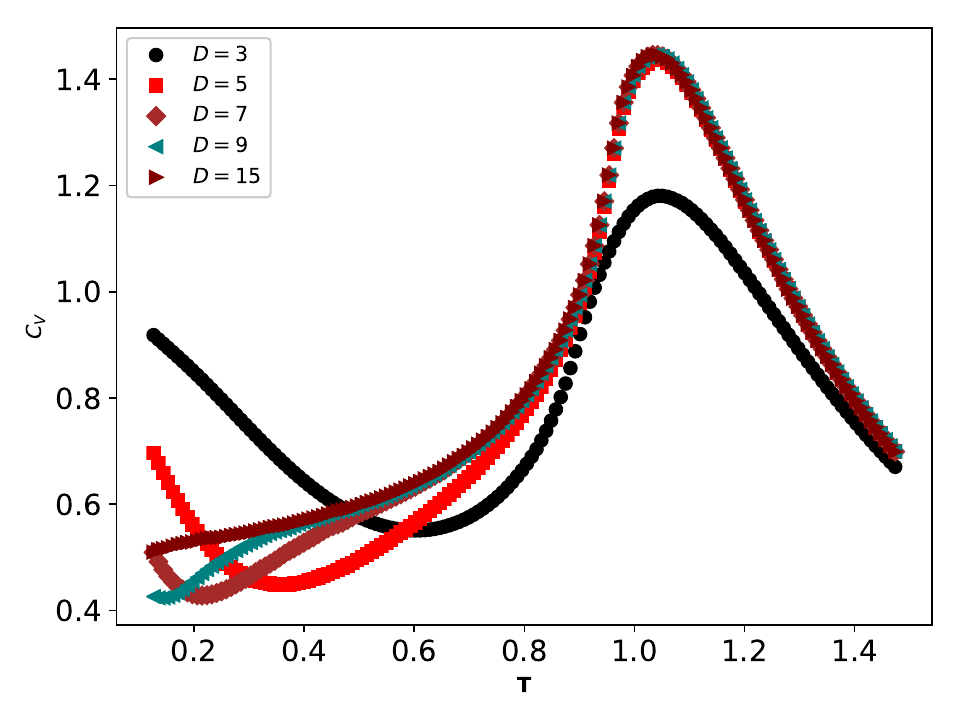}
  \caption{Specific heat, $C_{V}$, versus temperature, $T$, for the truncated XY model with varying initial bond dimension $D$.}
  \label{fig:cvxy}
 \end{minipage}
 \hfill
 \begin{minipage}{0.48\textwidth}
  \includegraphics[width=\textwidth]{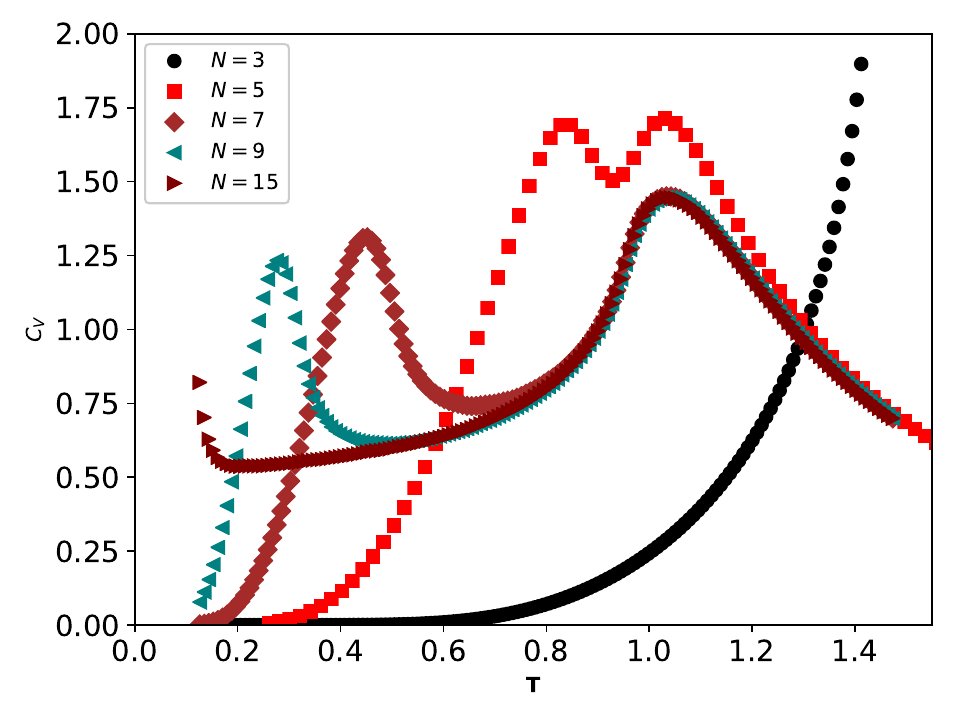}
  \caption{Specific heat, $C_{V}$, versus temperature, $T$, for the $Z_N$ clock model with varying $N$.}
  \label{fig:cvn}
 \end{minipage}
\end{figure*}

\section{Discrete approximations of continuous symmetry}
\label{sec:disc}

Here we compare different ways to approximate the continuous XY model
with a fixed number of states in the tensor network representation.
In particular we compare the truncation of the XY model expanded in the
character basis \cite{Liu:2013nsa}
to the $Z_N$ clock model for the same number of states.

We first look at the effect of truncation on the
specific heat, $C_V = -T \partial_T^2 F$,
versus temperature, $T$, with varying number of states.
Here, $F = - (\ln Z)/(\beta V)$ is the free energy.
While the specific heat does not exhibit critical behavior in the
2D XY model, and is therefore not a reliable indicator of the location
of the phase transitions,
it still serves as a qualitative indicator of the phase structure of the
theory.
For this purpose we obtained the specific heat results at $D_{cut}=40$,
from the second derivative of a seven-point polynomial fit to the free energy.
We will also consider another observable that is critical below.

In the case of the XY model, Fig.~\ref{fig:cvxy},
for all values of the initial bond dimension $D$,
we see a peak in $C_V$ around $T \approx 1$ which is larger than the 
BKT transition temperature of $T_c^{XY} \approx 0.89$
\cite{hasenbusch2005two,PhysRevE.89.013308,jha2020critical,ueda2021resolving}.
The height of the $D=3$ peak is very different from the rest ($D\geq 5$).
We also see $C_V$ rise again at low temperature.
This is more pronounced at small $D$ and the effect moves
to lower temperatures for larger $D$.
At $D=15$, the specific heat matches that from larger $D$
(up to $D=39$) within $0.1\%$ difference down to $T=0.2$.

We can compare the results from the truncated XY model to those of the clock models for the same number of states.
In Fig. \ref{fig:cvn} we show the specific heat versus temperature for the $N$-state clock models over the same range of $N$ as for $D$ before.
Again we see that for three states, the behavior deviates from the rest, but it is even more enhanced in this case.
For $N \ge 5$ the model develops a clear second peak in the specific heat
at lower temperature.
The $N=5$ specific heat is closer to $N \ge 7$ than for $N=3$,
but still deviates fairly significantly.
For $N\ge 7$ the peak around $T=1$ remains consistent among the models
with the main difference being the lower temperature peak moving towards
$T=0$ as $N$ increases.
The deviations at lower temperature are larger for the same
number of states as for the truncated XY case.

\begin{figure*}[htb!]
 \centering
 \begin{minipage}{0.48\textwidth}
  \includegraphics[width=\textwidth]{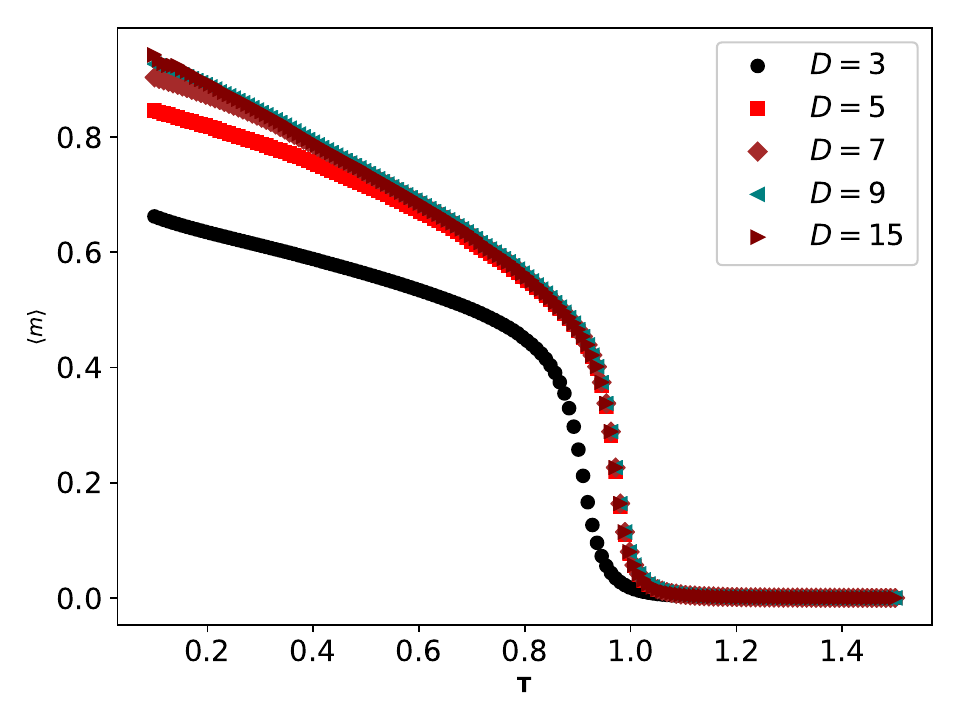}
  \caption{Magnetization per site, $\langle m \rangle$, versus temperature, $T$, for the truncated XY model with magnetic field, $h=10^{-4}$, and varying initial bond dimension $D$.}
  \label{fig:md}
 \end{minipage}
 \hfill
 \begin{minipage}{0.48\textwidth}
  \includegraphics[width=\textwidth]{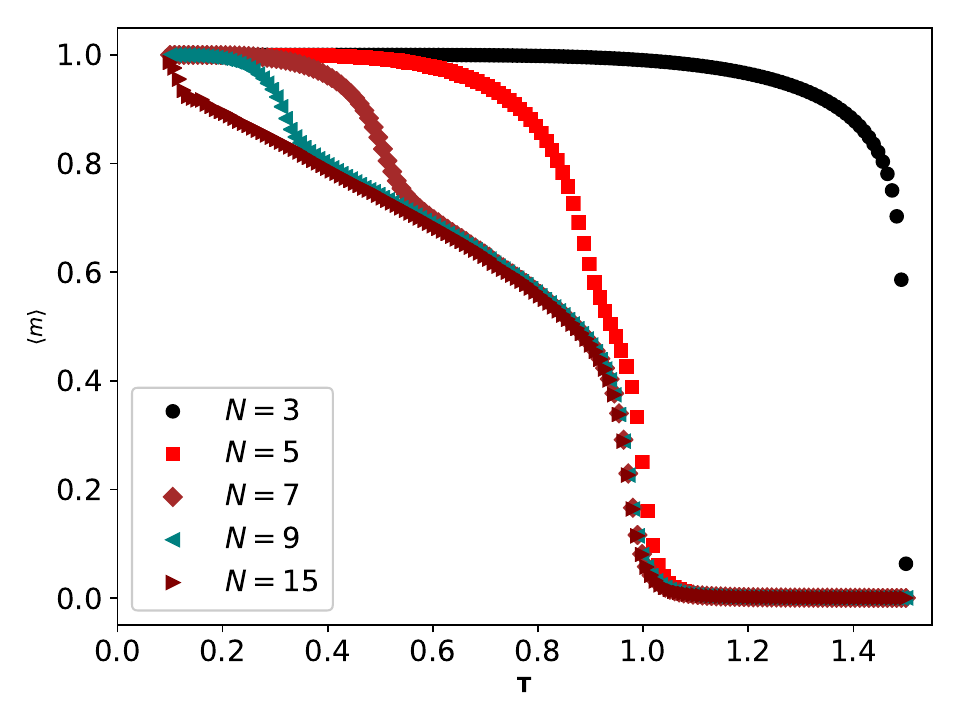}
  \caption{Magnetization per site, $\langle m \rangle$, versus temperature, $T$, for the $Z_{N}$ clock model with magnetic field, $h=10^{-4}$, and varying $N$.}
  \label{fig:magz}
 \end{minipage}
\end{figure*}

In Fig.~\ref{fig:md} and Fig.~\ref{fig:magz} we compare the behavior of the average magnetization per site,
$\langle m \rangle = \partial_h F$,
with a small magnetic field, $h=10^{-4}$,
between the two discretization schemes
for the same number of initial states. 
In the case of the truncated $XY$ model,
the magnetization never saturates to unity as we lower the temperature.
However in the $Z_N$ clock models,
we observe saturation at an intermediate temperature.
As $N$ increases, the saturation to unity happens at lower temperatures.

\begin{figure*}[htb!]
\centering
 \begin{minipage}{0.48\textwidth}
  \includegraphics[width=\textwidth]{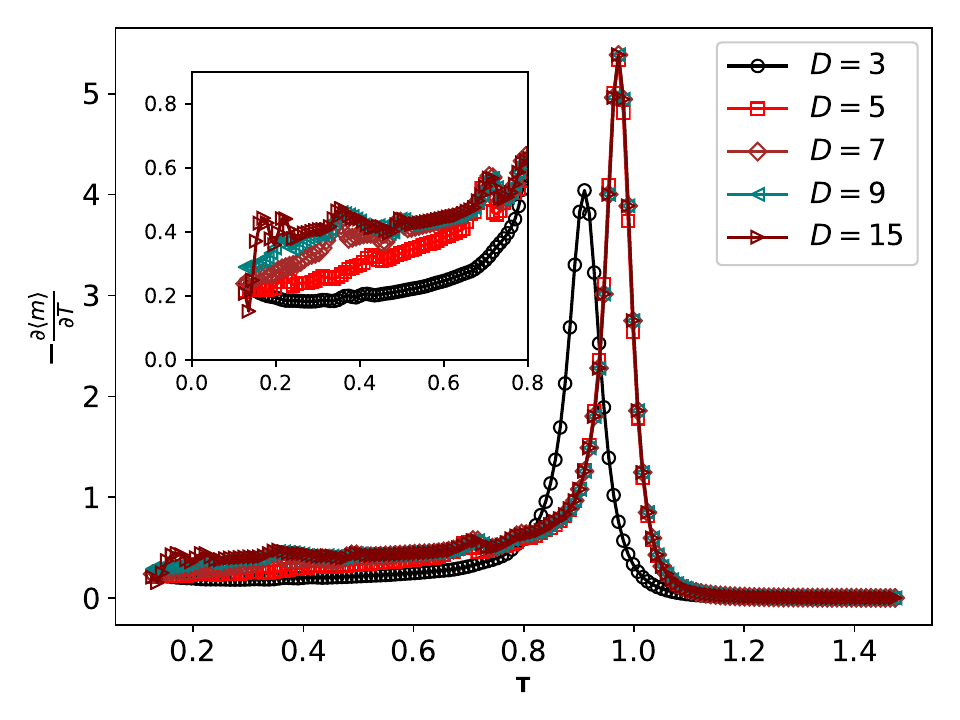}
  \caption{Cross-derivative, $-\partial_T \langle m \rangle$, versus temperature, $T$, for 
  the truncated XY model with $h=10^{-4}$ and varying $D$.}.
  \label{fig:dmdtd}
 \end{minipage}
 \hfill
 \begin{minipage}{0.48\textwidth}
  \includegraphics[width=\textwidth]{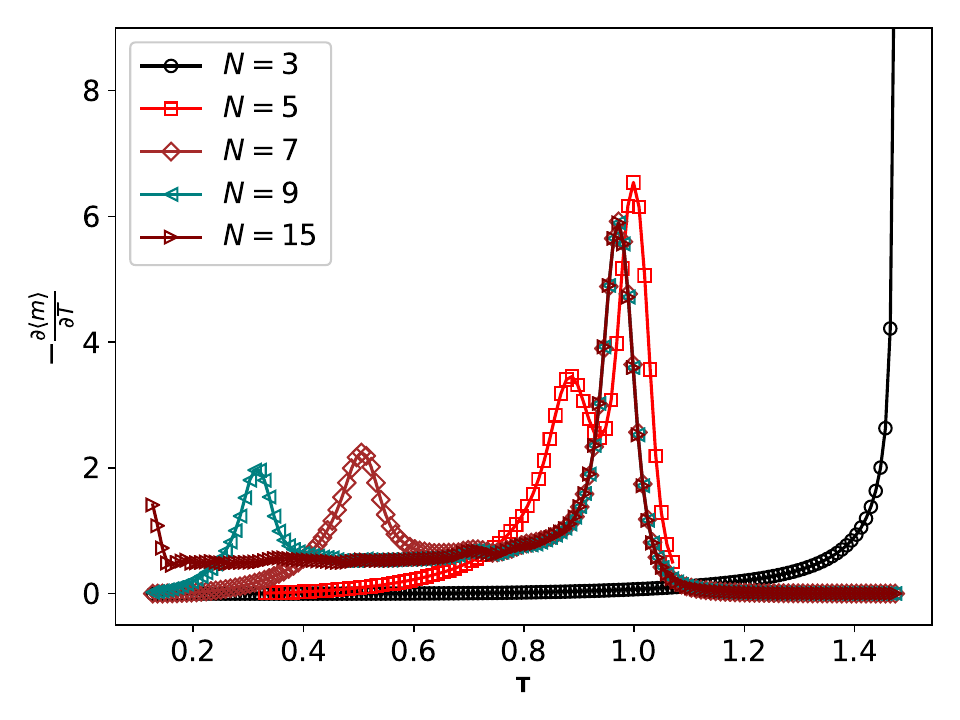}
  \caption{Cross-derivative, $-\partial_T \langle m \rangle$, versus temperature, $T$, for 
  the $Z_N$ clock models with $h=10^{-4}$ and varying $N$. }
  \label{fig:dmdT}
 \end{minipage}
\end{figure*}

It is difficult to observe the lower temperature transition using
magnetic susceptibility.
Instead we use a cross-derivative
\be
\frac{\partial^{2} F}{\partial h \partial T }= - \frac{\partial \langle m \rangle}{\partial T}
\ee
introduced in \cite{PhysRevB.101.165123}
to locate the lower temperature transition in the $Z_N$ models.

The cross-derivative of the magnetization for the XY model,
Fig.~\ref{fig:dmdtd}, shows only a single peak for all values
of the initial number of states, as low as $D=3$.
Here the location of the $D=3$ peak differs from $D \ge 5$, in contrast to the 
specific heat.
This is comparable to results seen in the $S=1$ quantum XY model ~\cite{PhysRevB.103.245137}.
For $D \ge 5$ the results are nearly identical for larger temperatures and only
show small changes in the value at lower temperatures (shown in inset).

For the $Z_N$ models, the cross-derivative of the magnetization,
Fig.~\ref{fig:dmdT}, clearly shows two peaks, indicating two phase transitions.
The peak height for the lower transition is smaller than the peak height for 
upper transition.
As $N$ is increased, we see that the lower peak moves towards lower temperatures. 
In this case it is clear that the truncated $XY$ model using the character basis
approaches the XY model behavior quicker than the discrete
$Z_N$ models for the same number of states.

\begin{figure*}[htb!]
 \centering
 \begin{minipage}{0.48\textwidth}
  \includegraphics[width=\textwidth]{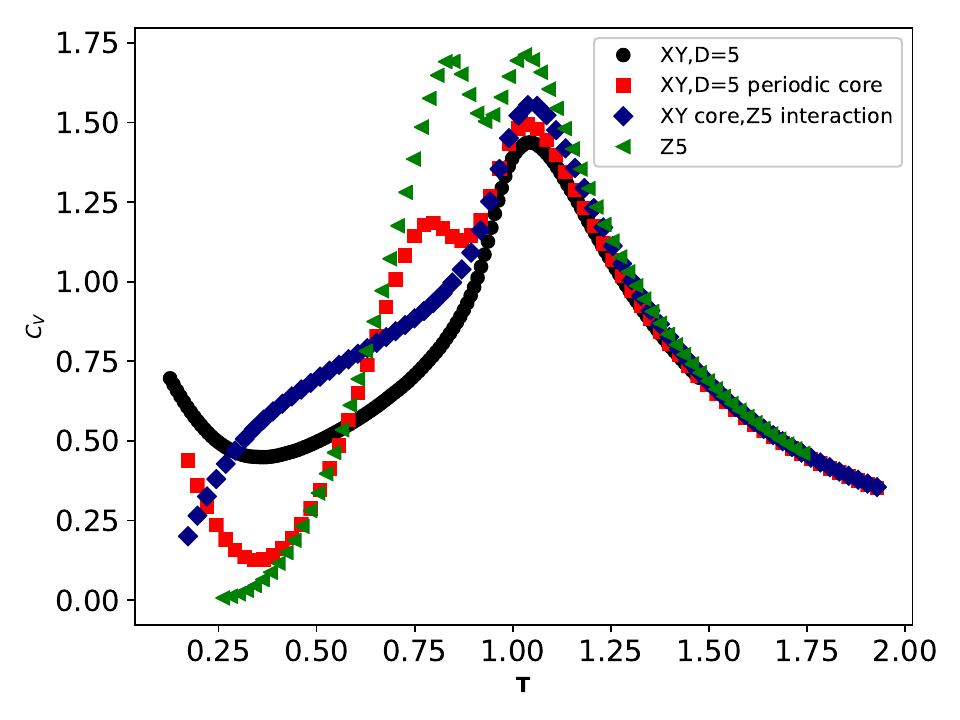}
  \caption{Specific heat, $C_V$, versus temperature, $T$, for different combinations of core tensor and interaction matrix from the 
  truncated $XY$ model at $D=5$ and the $Z_5$ clock model.
  The combination with the XY interaction matrix and $Z_5$ (periodic) core
  (squares) shows two peaks, similar to the $Z_5$ model (triangles),
  while the $Z_5$ interaction and $XY$ core (diamonds) shows no sign of a
  second phase transition.}
  \label{fig:xyvszncore}
 \end{minipage}
 \hfill
 \begin{minipage}{0.48\textwidth}
  \includegraphics[width=\textwidth]{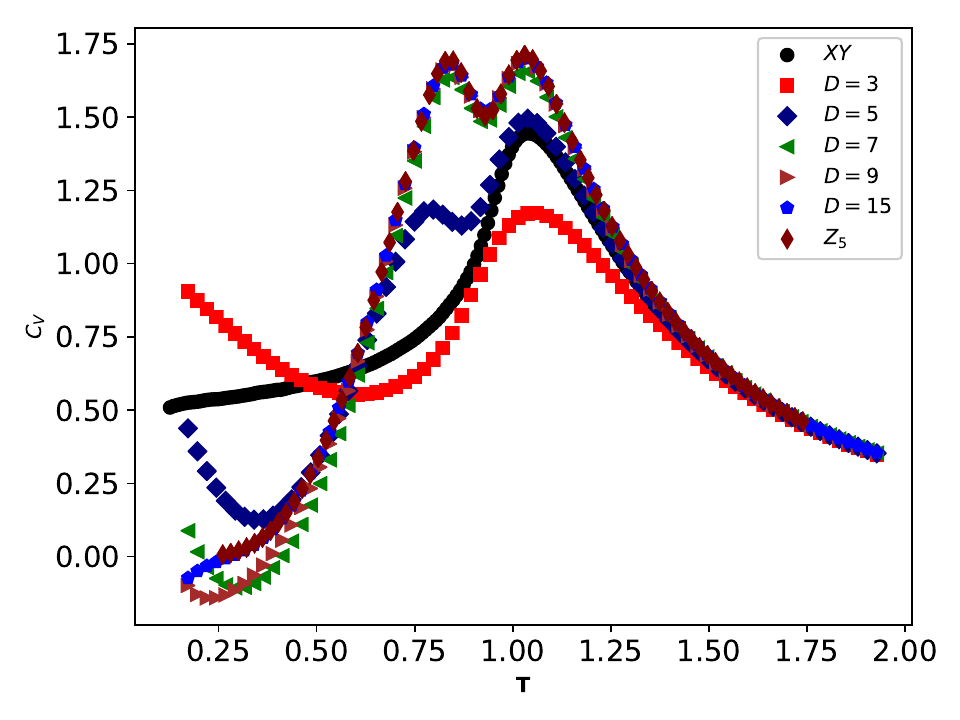}
\caption{Specific heat, $C_{V}$, versus temperature, $T$, for a
mixed model with the XY interaction matrix with a $Z_5$ core tensor
generalized to different numbers of initial states $D$.
For $D=5$ this corresponds to the same mixed model in 
Fig. \ref{fig:xyvszncore} (squares).
For large $D$ this approaches the $Z_5$ clock model result (diamonds).
The $XY$ model (circles) is also shown for comparison.}
\label{fig:z5vsxy5}
 \end{minipage}
\end{figure*}

\section{Role of core tensor and interaction matrix}
\label{sec:role}

While both the core tensor and the interaction matrix
differ between the XY and clock models, it is expected that
the core tensor, which implements the selection rules responsible for
charge conservation at $h=0$, is mainly responsible
for determining the symmetries of the theory \cite{PhysRevD.100.014506}.
We can check this by combining the core tensor and interaction matrix from 
the different theories and compare them to the original models.

In Fig.~\ref{fig:xyvszncore} we compare the different combinations
of the core tensor and interaction matrix from the $Z_5$ clock model
and the XY model at $D=5$.
The combination with the XY interaction matrix and $Z_5$ (periodic) 
core tensor clearly shows two peaks.
This is similar to the $Z_5$ model, though the lower temperature peak
in the mixed case is much smaller.
The mixed model with a XY core and $Z_5$ interaction shows only
a single peak within the calculated temperature range, similar
to the XY model.
For these models it is clear that the structure of the core tensor
is the determining factor in the overall phase structure of the theory.

In Fig.~\ref{fig:z5vsxy5} we plot the specific heat of the $XY$ model
with different numbers of initial states and a suitably generalized
$Z_5$ periodic core tensor.
For a large number of initial states the model reproduces the
specific heat of $Z_5$ clock model, and approaches
the infinite dimensional representation of the $Z_N$ models
shown in eqs. (\ref{CN}) and (\ref{IN}).
For $D=3$ there is only one peak, while for $D \ge 5$ there are two peaks.
The lack of a second peak in the $D=3$ case can be understood
due to there not being enough states to fully feel the effects of the
$Z_5$ periodicity in the core tensor.
This representation of the clock models provides a basis
for studying the transition between them and the XY model.

\section {Perturbed XY Model}
\label{sec:pert}

In order to study the transition from the XY to clock models,
we consider a perturbed model by adding an additional term, 
eq. \eqref{pxy}, which has an exact $Z_N$ symmetry.
This is equivalent to the model studied by 
Jos\'e, Kadanoff, Kirkpatrick, and Nelson (JKKN) \cite{PhysRevB.16.1217,PhysRevB.17.1477}.
We choose a normalization, eq. \eqref{pxynorm}, for this model such that
the $h_N \to \infty$ limit reproduces the $Z_N$ clock model.
For numerical purposes, we use an approximation to the
core tensor in eq. \eqref{CPXY}.
By writing it in the alternate form
\be
C^{XYN}_{abcd} &=& \sum_{k\ell} I_k(\beta h) 
\frac{I_\ell(\beta h_N)}{I_0(\beta h_N)} \delta_{a-b+c-d+k+N\ell} ~,
\ee
we can evaluate this for small $h$ by limiting the sum to
$-2 \leq k \leq 2$.
This form is accurate up to $O(h^2)$, which is
sufficient for our purposes.

For this model, we will always set $D = D_{cut}$, so that we are no longer comparing the effects of truncation on the phase diagram.
Instead we are interested in looking at the effects of a 
symmetry breaking perturbation on the continuous symmetry,
separate from the effects of the truncation.
One could also combine the two and consider smaller truncation of the perturbed
XY model, but for simplicity, we do not consider that here.
We note however, that the truncated XY model at $D=5$ is already very close
to the large $D$ limit of the XY model (see e.g. Fig. \ref{fig:dmdtd}),
so we expect the main results here to carry over even down to $D=5$.

We simulated this model at $N=5$, which is the smallest value of $N$ for which a second phase transition appears in the clock model.
We are interested in seeing how the extra phase transition emerges for small $h_5$.
The HOTRG method used here requires an increasingly large $D_{cut}$ to get stable results as we approach the continuous XY model,
which limits the lower value of $h_5$ which we could confidently
simulate with $D_{cut}=91$.
In lieu of direct simulations at very small $h_5$ we instead must 
extrapolate our results to small $h_5$.

The choice of $D_{cut}=91$ for the bulk of the simulations in this section
was made from a comparison of selected results for a range of $D_{cut}$ values.
$91$ was determined to be the smallest value that had acceptable errors for
the quantities measured.
A more detailed comparison of the errors with varying $D_{cut}$
is given in Appendix \ref{sec:dcut}.

\begin{figure*}[htb!]
 \centering
 \begin{minipage}{0.48\textwidth}
  \includegraphics[width=\textwidth]{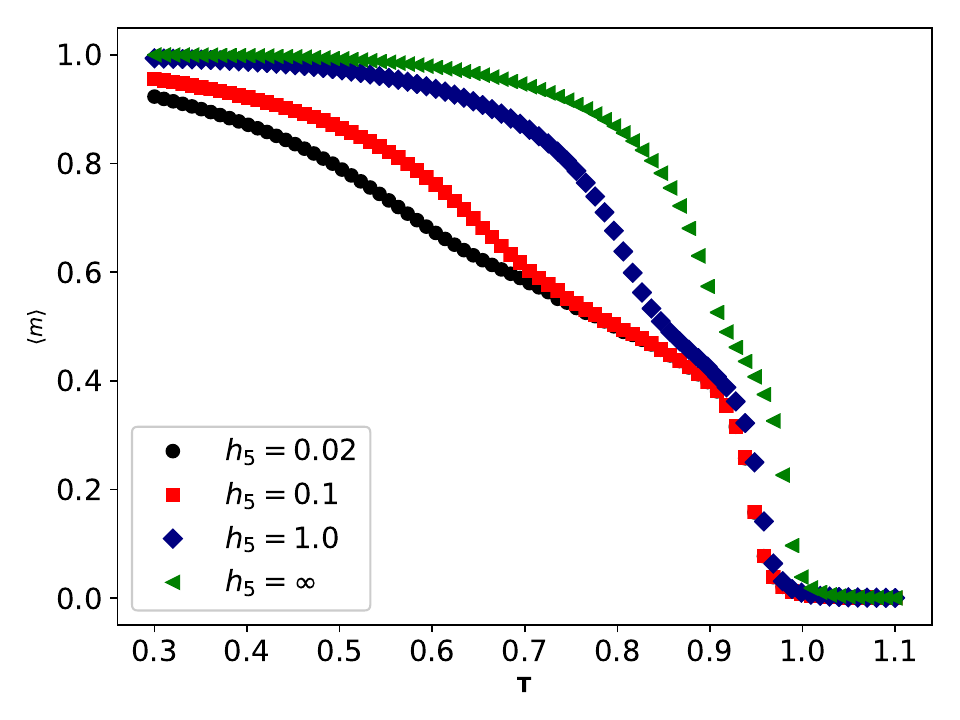}
  \caption{Magnetization per site, $\langle m \rangle$, versus temperature,
  $T$, for the perturbed XY model with magnetic field, $h=10^{-5}$, and varying $h_{5}$.}
  \label{fig:magh5}
 \end{minipage}
 \hfill
 \begin{minipage}{0.48\textwidth}
  \includegraphics[width=\textwidth]{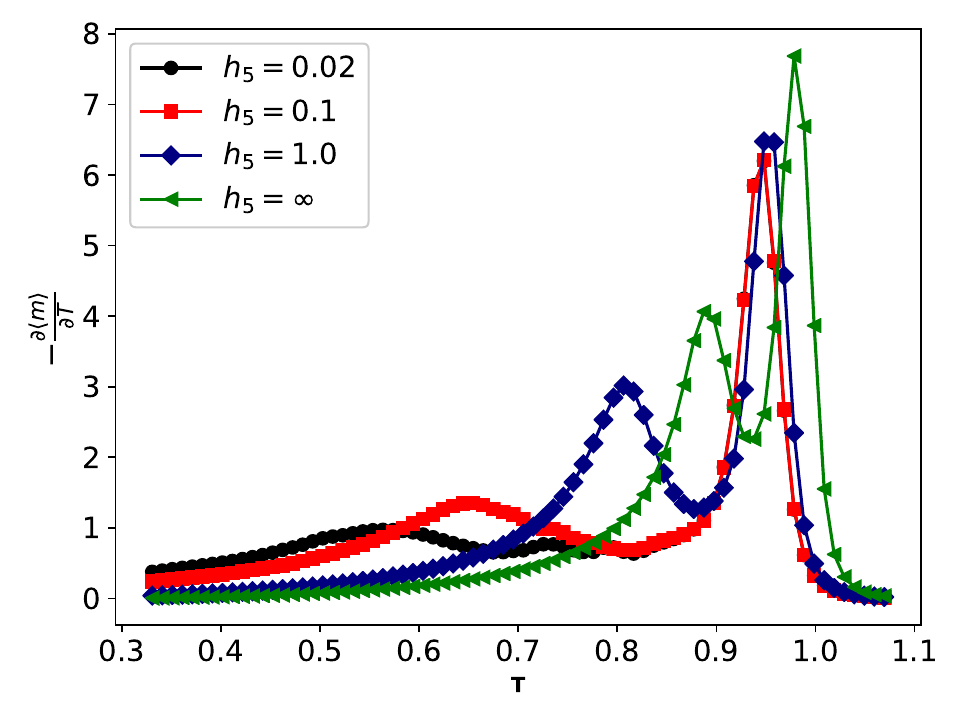}
  \caption{Cross-derivative $-\partial_T \langle m \rangle$ versus temperature, $T$, for the perturbed XY model with magnetic field, $h=10^{-5}$, and varying $h_5$.} 
  \label{fig:dmdth5}
 \end{minipage}
\end{figure*}

In Fig.~\ref{fig:magh5} we plot the magnetization versus temperature with a small magnetic field of $h=10^{-5}$ for different values of $h_5$ produced from HOTRG simulations with $D_{cut}=40$.
The magnetization profile smoothly interpolates between 
the $XY$ and $Z_5$ models as the symmetry breaking field, $h_5$,
is varied in a continuous manner.

Again we will use the temperature derivative of the magnetization \cite{PhysRevB.101.165123} to more clearly identify the phase
transitions.
The temperature derivative is obtained numerically from 
local polynomial fits to the magnetization.
In Fig. \ref{fig:dmdth5} we show $\dmdt$ versus temperature for a range
of $h_5$, again with $h=10^{-5}$.
Here we clearly see two peaks for all values of $h_5$ shown.
The upper peak occurs at a temperature, $T_{p2}$, which shifts a little
between $h_5=\infty$ (the $Z_5$ clock model) and $h_5=1$,
but is fairly stable below that.
In contrast, the temperature of the lower peak, $T_{p1}$,
continues to move towards
lower temperatures, and also the peak height decreases.
We will see below, however, that $T_{p1}$ does not seem to go to zero as $h_5 \to 0$,
and is consistent with having a phase transition at a temperature
that is significantly away from zero as $h \to 0$,
for all values of $h_5$.

We obtain values for $T_{p1}$ and the corresponding peak height from a
series of fourth order polynomial fits to the magnetization versus temperature using an interval of 11 data points, which represent a
range in temperatures of $\delta T = 0.04$.
This range was chosen as a good compromise between smoothing out any
small fluctuations in the data while preserving the sharpness of the
peaks as much as possible.
Values of $T_{p1}$ are obtained from the zeros of the second
derivative of the fitted polynomials.
Multiple values of $T_{p1}$ and the height are obtained by shifting
the fit interval, and the values and estimated errors are taken from the average and variance among the intervals that contain the peak.

\begin{figure}
 \includegraphics[width=\linewidth]{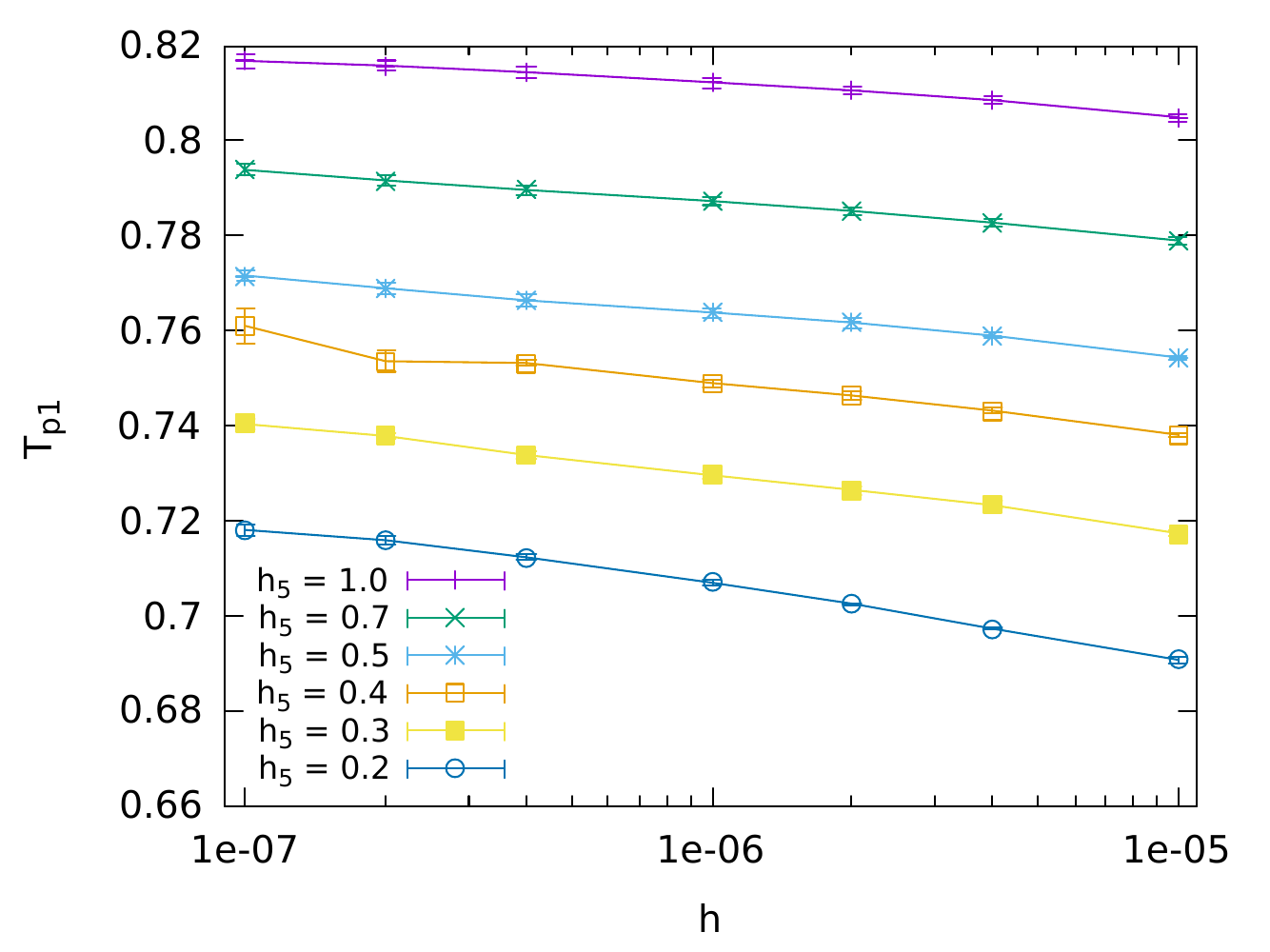}
 \caption{Lower peak temperature, $T_{p1}$, versus magnetic field, $h$,
 for the perturbed XY model with varying $h_{5}$.}
 \label{fig:ptvh}
\end{figure}

The results for $T_{p1}$ for a range of $h$ and $h_5$ values are 
shown in Fig. \ref{fig:ptvh}.
All these results were obtained using a bond dimension of $D=91$.
As noted above, the HOTRG results become less consistent as we approach
smaller values of $h$ and $h_5$.
While we could get stable results for much smaller $h_5$
within the range $10^{-7} \le h \le 10^{-5}$,
we could only confidently extract accurate peaks for $h_5$ down to 0.2.
For $h_5 < 0.2 $ the data becomes less smooth so that extracting accurate peaks becomes difficult.

\begin{table}[htb]
\centering
\begin{tabular}{ |c|c|c| } 
 \hline
 $h_5$ & $T_{c1}$ & $\gamma$ \\ 
 \hline
 0.2 & 0.750(9) & 0.1257(12) \\
 0.3 & 0.801(39) & 0.1195(19) \\
 0.4 & 0.787(24) & 0.1336(20) \\
 0.5 & 0.791(9) & 0.1295(65) \\
 0.7 & 0.812(6) & 0.1315(28) \\
 1.0 & 0.823(1) & 0.1353(25) \\
 $\infty$ & 0.907(3) & 0.1220(51) \\
 \hline
\end{tabular}
\caption{Table of fit results for the lower critical temperatures, $T_{c1}$,
from eq. \eqref{tc1}, and peak height scaling exponent, $\gamma$, 
from eq. \eqref{gamma}, versus $h_5$.
$h_5=\infty$ corresponds to the $Z_5$ clock model.
The errors are statistical for $D_{cut}=91$. Systematic errors due to finite $D_{cut}$ are discussed in the text and in 
Appendix \ref{sec:dcut}.}
\label{fitdata}
\end{table}

For each value of $h_5$, we fit the data over the range
$10^{-7} \le h \le 10^{-5}$
to the power-law scaling formula
\be
\label{tc1}
T_{c1} - T_{p1} \propto h^{\alpha} ~.
\ee
The fit values of the critical temperature, $T_{c1}$, are summarized
in Table \ref{fitdata}.
Unfortunately, the error bars are too large to reliably extrapolate this
critical temperature to $h_5=0$.

\begin{figure}
 \includegraphics[width=\linewidth]{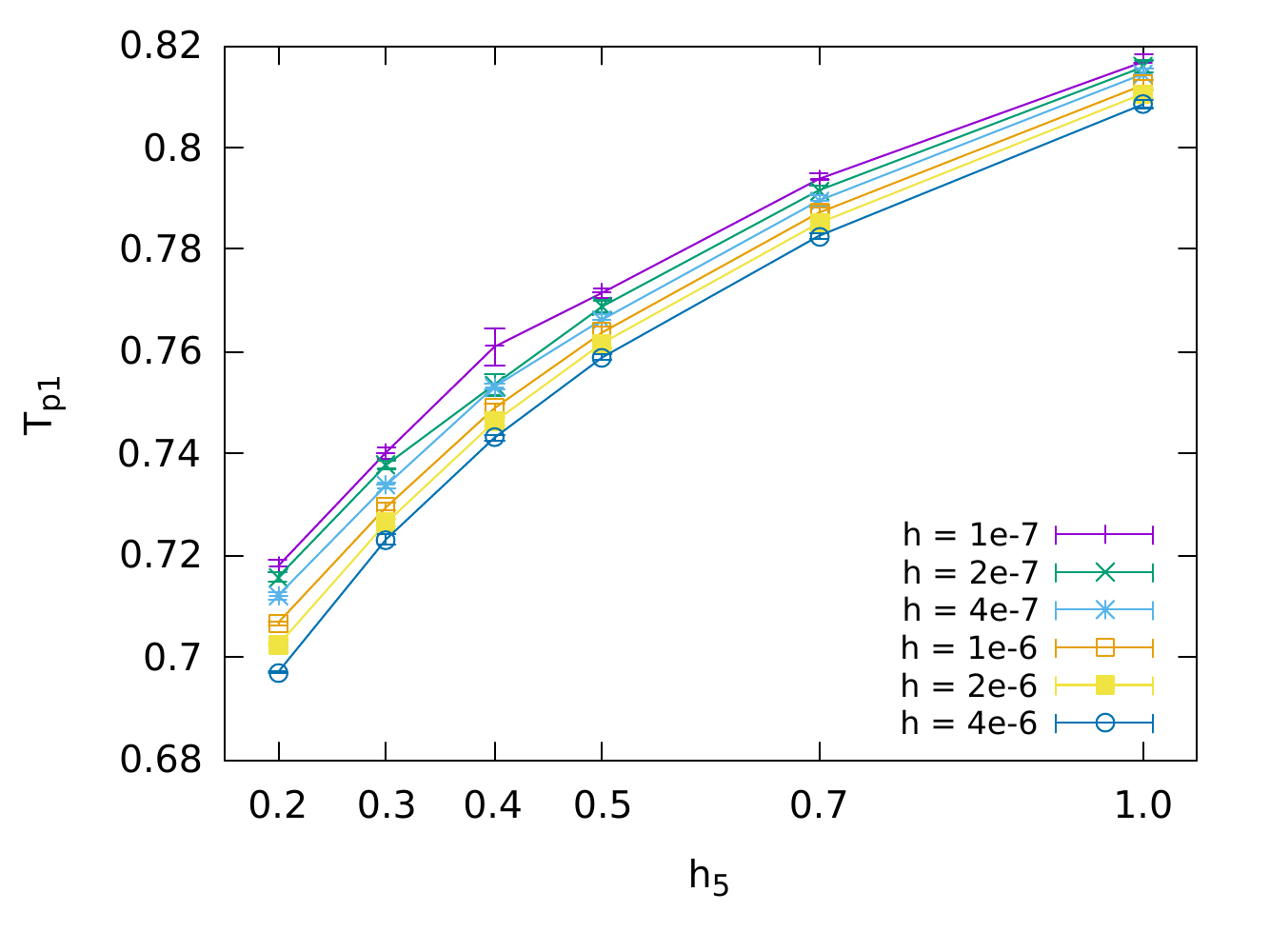}
 \caption{Lower peak temperature, $T_{p1}$, versus $h_5$ for the perturbed XY model with varying $h$.}
 \label{fig:ptvhn}
\end{figure}

An alternate approach is to first extrapolate $h_5 \to 0$, then take the $h \to 0$ limit.
In Fig. \ref{fig:ptvhn} we use the same data from Fig. \ref{fig:ptvh} to plot
$T_{p1}$ versus $h_5$ for different values of $h$.
We fit the data at fixed $h$ over the range $0.2 \le h_5 \le 1.0$
to another power-law form
\be
T_{p1} - T^*_{p1} \propto h_5^{a} ~,
\ee
where $T^*_{p1}$ is the fit result for the lower peak temperature extrapolated
$h_5 \to 0$.
These fits show good agreement across the whole range in $h_5$.

\begin{figure}
 \includegraphics[width=\linewidth]{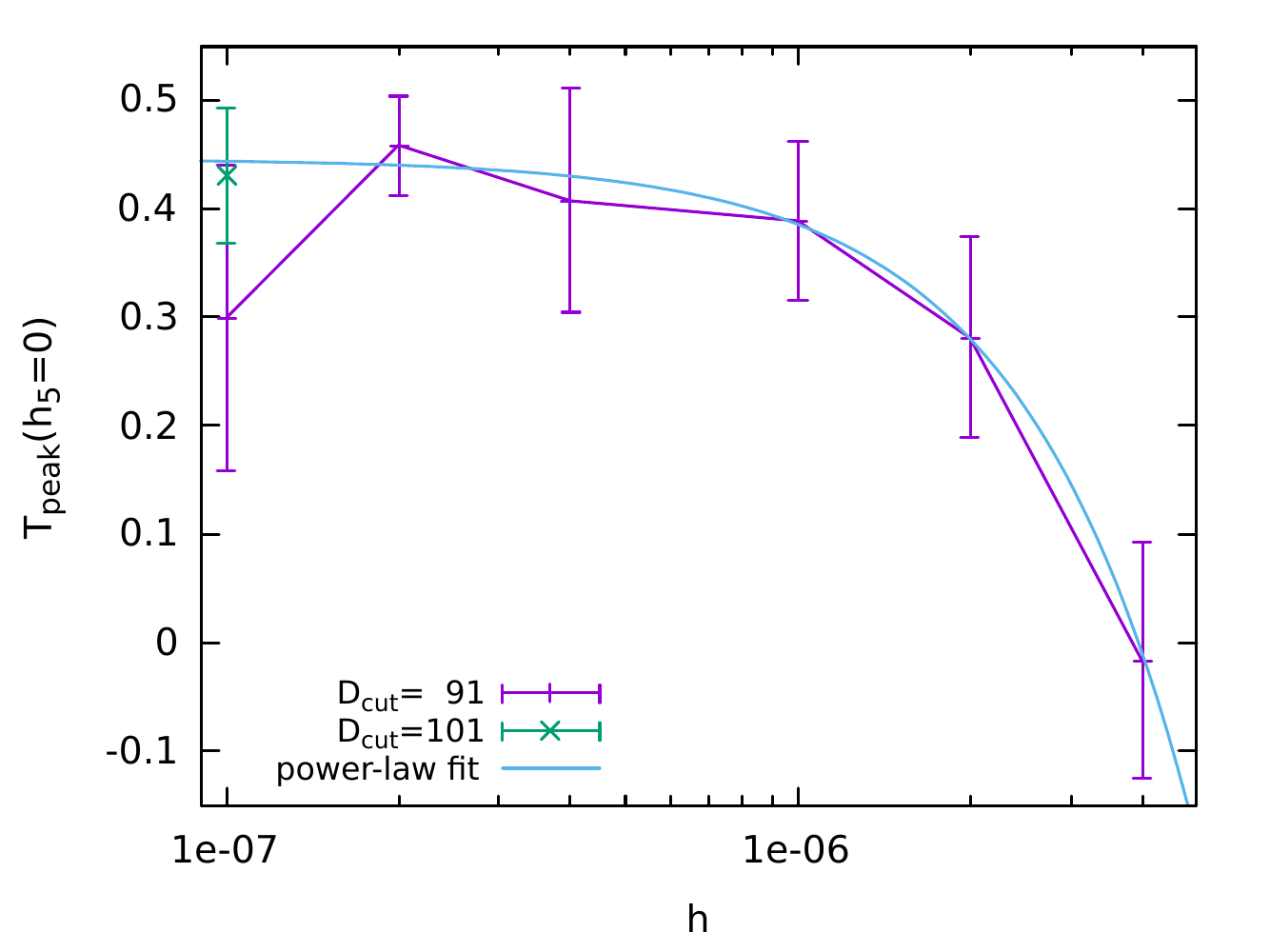}
 \caption{Lower peak temperature extrapolated to $h_5=0$
 ($T_{p1}^*$ in the text) versus magnetic field, $h$, for the perturbed XY model.
 The smooth line is a fit to the power-law form in eq. \eqref{tc1star}.
 The smallest $h$ value also has a result for $D_{cut}=101$, which is consistent with
 the $D_{cut}=91$ results.}
 \label{fig:ptsvh}
\end{figure}

The values of the extrapolated peak temperatures, $T_{p1}^*$,
in the limit $h_5 \to 0$ are plotted in Fig. \ref{fig:ptsvh}.
For larger $h$,
the extrapolated peak temperature is consistent with zero.
However for smaller $h$, the value of $T_{p1}^*$ moves away from zero.
A power-law fit to the form
\be
\label{tc1star}
T^*_{c1} - T^*_{p1} \propto h^{\beta} ~,
\ee
where $T^*_{c1}$ is the fit result for the lower peak temperature extrapolated
$h_5 \to 0$ and $h \to 0$, is shown along with the data.
The extracted value of the critical temperature at $h=0$ and $h_5=0$ is
$T^*_{c1} = 0.44(3)$, which is about half of the XY $T_c$ of about $0.89$.
A similar result was obtained even when leaving out the highest
$h=4 \times 10^{-6}$ point,
so that point is not a significant factor in
constraining the curvature.
We also have a data point for $D_{cut}=101$ at $h=10^{-7}$ included in the
plot.  This point is nicely consistent with the $D_{cut}=91$ point and
the power-law fit.

\begin{figure}
 \includegraphics[width=\linewidth]{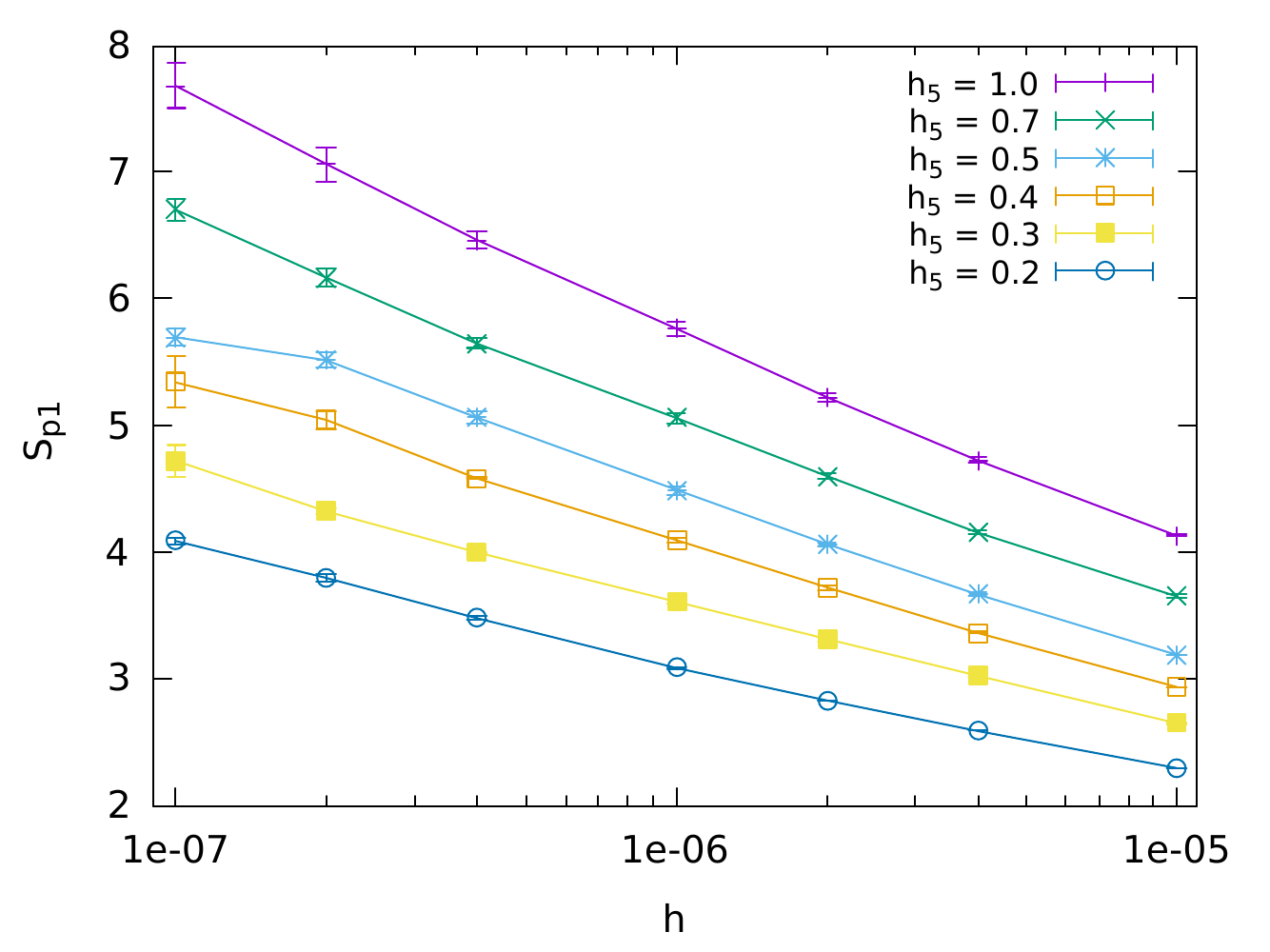}
 \caption{Lower peak height, $S_{p1}$, versus magnetic field, $h$,
 for the perturbed XY model for varying $h_{5}$.}
 \label{fig:phvh}
\end{figure}

In order to establish whether the lower peak in the cross-derivative,
$\dmdt$, corresponds to a true phase transition,
we look to see if the scaling of the peak height exhibits a critical behavior.
In Fig. \ref{fig:phvh} we plot the lower peak height, $S_{p1}$,
versus $h$ for different $h_5$.
We see that the heights do increase as $h \to 0$.
We fit the peak height in the range $h \in [1e-7,4e-6]$ to a power-law scaling form
\be
\label{gamma}
S_{p1} \propto h^{-\gamma} ~,
\ee
The extracted exponents, $\gamma$, calculated at $D_{cut}=91$
are listed in Table \ref{fitdata}.
The values for the exponent listed there, are clearly show that the
peak of the cross derivative is critical for the given values of $h_5$,
however there is still a significant systematic error due to the
fixed $D_{cut}$.
To get an estimate of this, we have also calculated
$\gamma=0.1191(14)$ at $h_5=0.2$ with $D_{cut}=101$.
This is lower than the corresponding value with $D_{cut}=91$,
but similar to that at $h_5=0.3$.
From the difference in the $D_{cut}=91$ and $101$ values we can
estimate that there is a systematic error of at least $0.0066$
that must be added to the above extrapolations.
If the systematic error is no more than a few times this lower estimate
then the exponents would still clearly favor a critical system.

To see that the exponent remains nonzero as $h_5$ goes to zero, we
performed some simple fits.
A linear fit to $\gamma$ for $h_5 \in [0.2,1.0]$ gives a value
of $\gamma = 0.1221(34)$ at $h_5 =0$,
while a quadratic fit yields $0.1202(97)$.
Both of these values are not consistent with zero, even when taking
into account the estimate of the systematic error considered above.
Thus the peaks appear to correspond to a critical phase transition
even as $h_5 \to 0$.
This result is in agreement with the RG analysis based conclusion
from \cite{PhysRevB.16.1217,PhysRevB.17.1477},
in which the $h_N \to 0$ limit corresponds to a phase
transition with non-zero critical temperature.

\section{Summary and Future Directions}

We have studied the changes in phase structure when moving
between the continuous XY model and the discrete $Z_N$ clock model
in two dimensions using the
Higher Order Tensor Renormalization Group approach.
We compared different ways to approximate the XY model
with a fixed number of states in the tensor network representation.
We showed 
that the truncated character expansion for the XY model 
approaches the continuous XY model behavior quicker than than discrete
$Z_N$ model with same number of states, as the number of states increases.

We also examined the role of the core tensor and interaction matrix
in determining the phase structure of the XY and $Z_N$ models,
and showed that the core tensor plays the dominant role in the
phase structure of these models.

Further we explored a perturbed XY model (the JKKN model)
which allows us to interpolate
between the XY and $Z_N$ models via the introduction of a symmetry
breaking term.
This term breaks the $U(1)$ symmetry to $Z_N$ and introduces a new phase transition
(for $N\ge 5$) in addition to the well-know BKT transition.
We demonstrate that the cross-derivative peak corresponding to this
phase transition in the lower temperature region scales with volume,
and that the critical temperature doesn't go to zero even in the limit of
a vanishingly small symmetry breaking field.
This suggests that even small symmetry-breaking perturbations 
can have a large effect on the phase structure of theories.
When discretizing models, and potentially when simulating them
in noisy environments, such as present quantum computers,
one must be careful to avoid these possible effects.

While the perturbation considered here had an explicit $Z_N$ symmetry,
which may have had a dominant effect on the resulting phase structure, it
would also be interesting to explore the effect of perturbations with
less symmetry to see how the results are modified.
For example, $N$ could be extended away from integer values as was considered
in the limit of a large perturbation here \cite{Hostetler:2021gsb,PhysRevD.104.054505}.
Determining the effect of small perturbations with varying degrees of
symmetry could be important to understanding the errors inherent
in simulations on resource limited and noisy near-term quantum
simulators.

In this paper we have studied the XY model due to its simplicity and the fact that
its group-space discretizations, the clock models, exhibit notably different
phase structure.
For future work it will be interesting to study the discrete truncations of other
continuous models such as the Heisenberg model,
O(N) vector model,
the compact hyperbolic spin model \cite{gubser2016nonlinear},
or gauge theories.
It would also be interesting to explore other expansion bases for discretization
to see their effects on the phase structure and if there are alternatives that could
produce better representations for a limited number of states.

\textbf{Acknowledgements}
N.B., X.-Y.J. and J.C.O. were supported through a QuantISED grant
by the U.S. Department of Energy, Office of Science, Office of High Energy Physics.
Z.S. is supported by the National Science Foundation under 
Award No. 2037984 and DOE Q-NEXT.
This research used resources of the Argonne Leadership Computing Facility,
which is a DOE Office of Science User Facility supported under Contract 
DE-AC02-06CH11357.

\appendix

\section{Tensor Network construction}
\label{sec:TNconstruction}

Here we give a general treatment on the construction of the tensor
network formulation of spin model partition functions.
This is a generalization of the method used for the XY model \cite{Liu:2013nsa}.

We start with a Hamiltonian that is split into a nearest neighbor
interaction term, $H_I$, and local term, $H_L$, as
\be
  H = \sum_{\langle ij\rangle} H_I(\theta_i,\theta_j)
    + \sum_{i} H_L(\theta_i)
\ee
where the $\theta_i$ are variables on a site of the lattice that
will be integrated over, and $\langle ij \rangle$ represents all pairs of neighboring sites.

Note that we can move local terms to or from the interaction term by defining
\be
\label{hiprime}
H_I'(\theta_i,\theta_j) &=&  H_I(\theta_i,\theta_j) 
+ \frac{1}{2d} \left[ F(\theta_i) + F(\theta_j) \right] \\
H_L'(\theta) &=& H_L(\theta) - F(\theta)
\ee
for arbitrary $F$.
Setting $F = H_L$ would then eliminate the local term and put the whole model
into the interaction term.
This will be a convenient form when discussing the $Z_N$ model below.
For the XY model we will consider the original form with the simpler interaction term.
When constructing the tensor network using a complete basis set, then choice of
convention for the local and interaction term does not matter.
However, when truncating the basis set, the different forms can give different results.

In constructing the tensor network formulation,
the main step is to separate the interacting variables by expanding the interaction term
in some set of basis functions,
$f_a(\theta)$ and $g_a(\theta)$, as
\be
\label{mdef}
\e^{- \beta H_I(\theta_i,\theta_j)} \approx
\sum_{ab} M_{ab} f_a(\theta_i) g_b(\theta_j)
\ee
with $M_{ab}$ a matrix characterizing the interaction.
After collecting terms with common integration variables we are
left with a tensor on each site.  On a two dimensional square lattice
this is given by
\be
C_{abcd} = \int d\theta ~ \e^{-\beta H_L(\theta)}
f_a(\theta) g_b(\theta)
f_c(\theta) g_d(\theta)
\ee
where $a,b$ and $c,d$ are pairs of opposing directions on the lattice.

The partition function is then a contraction of neighboring core tensors,
$C$, with the interaction matrices, $M$, placed in between.
For convenience, one typically defines a new tensor which incorporates the
interactions into the core to make evaluations simpler.
One possibility is to take the SVD of the interaction matrix as
$M = U \Lambda V$ and then combine the factors $\tilde{U} = U \sqrt{\Lambda}$
and $\tilde{V} = \sqrt{\Lambda} V$ with $C$ to get
\be
T_{abcd} = C_{ijkl} \tilde{U}_{ia} \tilde{V}_{bj}
\tilde{U}_{kc} \tilde{V}_{dl}  ~.
\ee
Note that this can be viewed as performing a change of basis functions
to make $M$ the identity.
This form is not unique though due to the {\em gauge freedom} in tensor
networks.
Performing a similarity transformation on opposing pairs
of indices of the tensor still produces the same partition function.

\subsection{XY model tensors}

A standard basis used for a $U(1)$ spin variables, related to the character expansion, is
\be
\label{fb}
f_a(\theta) &=& \e^{i a \theta} \\
g_a(\theta) &=& \e^{-i a \theta} ~.
\label{fb2}
\ee
If the interaction Hamiltonian is a function of $\theta_i-\theta_j$, then
this basis makes $M$ diagonal.
For the standard XY model with 
$H_I(\theta_i,\theta_j) = -\cos(\theta_i-\theta_j)$,
the interaction matrix in the Fourier basis above, eqs. \eqref{fb} and \eqref{fb2}, is
\be
\label{xym}
M^{XY}_{ab} = \delta_{ab} I_a(\beta) ~.
\ee

The core tensor for the XY model with a magnetic field and perturbation term
\be
H_L(\theta) = - h \cos(\theta) - h_N \cos(N \theta)
\ee
is given by
\be
C^{XYN}_{abcd} &=& \mathcal{N}
\int d\theta \, \e^{\beta h \cos(\theta) + \beta h_N \cos(N\theta)
+ i (a-b+c-d) \theta} ~~~~~~~~ \\
&=& 2 \pi \mathcal{N} \hspace{-1.5mm} \sum_{k,\ell=-\infty}^{\infty} \hspace{-1.5mm} I_k(\beta h) I_\ell(\beta h_N) \delta_{a-b+c-d+k+N\ell} \\
&=& 2 \pi \mathcal{N} \sum_{\ell=-\infty}^{\infty} I_{a-b+c-d+N\ell}(\beta h) I_\ell(\beta h_N) ~.
\ee
In order to match the $Z_N$ model (given below) in the large $h_N$ limit,
we have included a normalization factor
\be
\mathcal{N} = \frac{1}{2 \pi I_0(\beta h_N)}  ~.
\label{pxynorm}
\ee
This scales the full partition function by a factor of $\mathcal{N}^V$.
Using the asymptotic form of the Bessel functions \cite{NIST:DLMF:10.30.E4}
\be
I_\ell (\beta h_N) \approx 
\frac{1}{\sqrt{2 \pi \beta h_N}} \e^{\beta h_N} ~,
\ee
which is independent of $\ell$, we have
\be
\lim_{h_N \to \infty} C^{XYN}_{abcd} &=&
 \sum_{\ell=-\infty}^\infty I_{a-b+c-d+N\ell}(\beta h) ~,
\ee
which, as we will show below,
is a form of the $Z_N$ clock model
written with an infinite basis set
(as opposed to the typical finite $N$-state basis).

\subsection{$Z_N$ clock model}

There are several ways to construct the tensor network for the $Z_N$ clock model.
Here we present a few variations starting with $N$-state 
representations in the discrete angle basis, and then in a character basis.
We also present an infinite state representation that coincides with
the $h_N \to \infty$ limit of the perturbed XY model.
Lastly we consider it as an approximation of the XY model using
a specific basis to expand the interaction term.

\subsubsection{$N$-state representations}

The simplest way to construct the $Z_N$ clock model tensor network is
to evaluate it in the basis of the discrete angles $\omega^N_a=2\pi a/N$.
This gives an interaction matrix of
\be
\label{znm0}
M_{ab} = \e^{ \beta \cos(\omega^N_a-\omega^N_b)}
\ee
and a core tensor of
\be
\label{znc0}
C_{abcd} = \e^{\beta h \cos(\omega^N_a)} \delta_{ab} \delta_{ac} \delta_{ad} ~.
\ee
The interaction matrix is not diagonal in this basis.
For comparison with the XY model, it is convenient to make it diagonal
by rotating to the character (Fourier) basis giving
\be
M^N_{ab} &=& \frac{1}{N^2} \sum_{k,\ell=-\infty}^\infty \e^{i (a \omega^N_k - b \omega^N_\ell) 
 + \beta \cos(\omega^N_k-\omega^N_\ell)} \\
&=& \delta_{ab} \sum_{k=-\infty}^\infty I_{a+kN}(\beta)  ~.
\ee
The core tensor in this basis becomes
\be
C^N_{abcd} &=& \frac{1}{N} \sum_{n=-\infty}^\infty 
\e^{i (a-b+c-d) \omega^N_n + \beta h \cos(\omega^N_n)} \\
&=& \sum_{\ell=-\infty}^\infty I_{a-b+c-d+N\ell}(\beta h) ~.
\label{zncore}
\ee
The normalizations in the above expressions were chosen for convenience.

\subsubsection{Infinite state representation}

One can get an infinite representation of the $Z_N$ clock model
by using the same basis, eqs. \eqref{fb} and \eqref{fb2},
and expansion of the interaction term as that used in the XY model.
This gives the same interaction matrix, eq. \eqref{xym}.
The core tensor is different than the XY model 
due to the discrete sum over angles.
Instead it takes the same form as the $Z_N$ core tensor
in eq. \eqref{zncore}, except that all the external indices,
$a$, $b$, $c$, and $d$, extend over the range $-\infty \ldots \infty$
instead of just over $N$ states.

The $N$-state $Z_N$ interaction matrix contains the same set of Bessel
functions as the infinite state one, just with them folded over on to
$N$ states.
This is a consequence of the periodicity of the core tensor
$C^N_{a+N,b,c,d} = C^N_{a,b,c,d}$, and similarly for other indices.

\subsubsection{Representation as approximation of XY model}

One can also view the $Z_N$ model as an approximation of the XY model
using the set of $N$ basis functions
\be
f_a(\theta) = g_a(\theta) = \chi^N_a(\theta)
\ee
with the rectangular step functions, $\chi^N_a(\theta)$, being
1 for $-\omega^N_{1/2} < \theta - \omega^N_a \le \omega^N_{1/2} \pmod{2 \pi} $
and 0 otherwise.
This basis divides the angles into $N$ disjoint parts, with each basis function
being constant over one part and zero on the others.

In this case we need to consider the
alternate form of the interaction term, $H_I'$ from eq. \eqref{hiprime}
with $F=H_L$, so that the local term is moved into the interaction.
The expansion of the interaction term, eq. \eqref{mdef},
can be approximated using
\be
M_{ab} = \e^{\beta \cos(\omega^N_a-\omega^N_b) 
+ (\beta h /4) \left[\cos(\omega^N_a) + \cos(\omega^N_b) \right]} ~.~~~~
\ee
The core matrix in this basis is simply a copy tensor
$C_{abcd} = \delta_{ab} \delta_{ac} \delta_{ad}$.
This form is similar to the original form given for the clock model
in eqs. \eqref{znm0} and \eqref{znc0}, and the interaction matrix
can be diagonalized using the same rotation to the character basis.

This basis provides a simple way to get a measure of the
amount of error in the truncation by examining the error
in the approximation of the interaction term.
One could compare the error in the expansion of the interaction term
among different finite basis choices.
This could then be used to choose a basis with the smallest error.

\section{Scaling with $D_{cut}$}
\label{sec:dcut}

\begin{figure}
 \includegraphics[width=\linewidth]{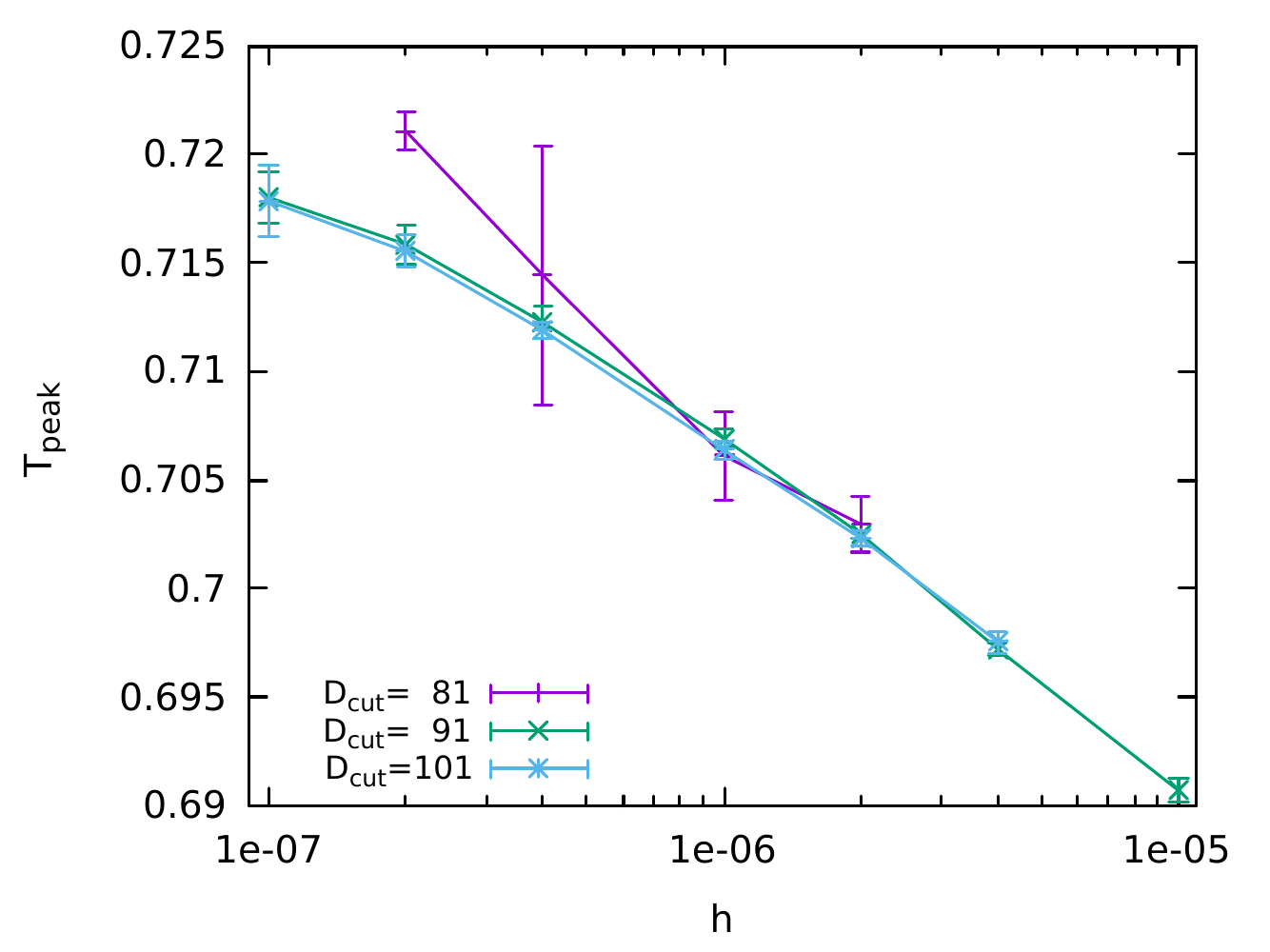}
 \caption{Lower peak temperature, $T_{peak}$, versus magnetic field, $h$,
 for the perturbed XY model at $h_5=0.2$ for varying $D_{cut}$.}
 \label{fig:ptvhdcut}
\end{figure}

To get an estimate of the errors due to a finite $D_{cut}$ in the
perturbed XY model, we performed calculations
with a range of $D_{cut}$ values for a few of the data points.
In Figure \ref{fig:ptvhdcut} we plot the fit value for the lower peak temperature
in the perturbed XY model with $h_5=0.2$ versus the magnetic field, $h$,
at different values of $D_{cut}$.
$D_{cut}=91$ and $101$ agree with each other within errors,
but $D_{cut}=81$ differs significantly at low $h$.
From this comparison we expect that $D_{cut}=91$ is sufficient
for extracting the peak temperature with error due to $D_{cut}$
smaller than the fit error.

\begin{figure}
 \includegraphics[width=\linewidth]{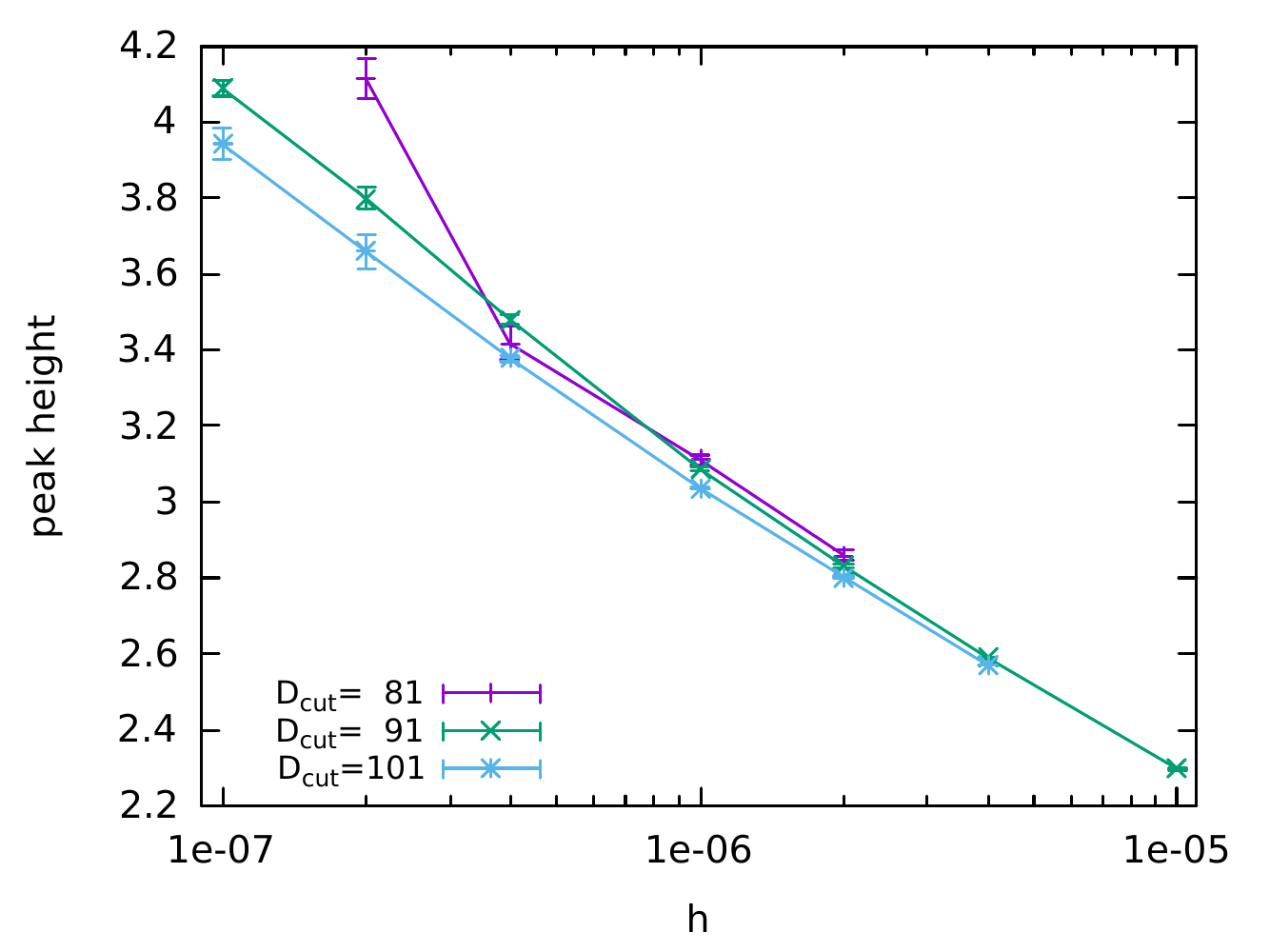}
 \caption{Lower peak height versus magnetic field, $h$,
 for the perturbed XY model at $h_5=0.2$ for varying $D_{cut}$.}
 \label{fig:phvhdcut}
\end{figure}

We also looked at the effect of $D_{cut}$ on the lower peak height,
shown in Figure \ref{fig:phvhdcut}.
Here again $D_{cut}=81$ varies significantly from the larger values
at smaller $h$.
However, here we see a discrepancy between $D_{cut}=91$ and $101$ too.
The peak height is lower in the $D_{cut}=101$ case, although the $91$ and $101$ curves are both smooth and have similar slopes on the semi-log plot.
The extracted exponent, $\gamma$, for $D_{cut}=91$ is $0.1257(12)$
(as reported in Table \ref{fitdata}), 
while the value for $D_{cut}=101$ is $0.1191(14)$.
As discussed in section \ref{sec:pert},
the difference between the $D_{cut}=91$ and $101$ values
gives a lower estimate for the systematic error.
Assuming the total systematic error is no more than a few times this
lower estimate, then the systematic error is not large enough
to make $\gamma$ consistent with zero.

\bibliography{ZN.bib}

\end{document}